\documentclass[preprintnumbers,superscriptaddress]{revtex4}

\usepackage{graphicx,bm,color,latexsym,amssymb}
\usepackage[latin1]{inputenc}

\definecolor{Blue}{rgb}{0.0,0.0,1}
\definecolor{Red}{rgb}{1,0.0,0.0}
\definecolor{Green}{rgb}{0,0.5,0.0}

\begin{document}

\title{Experimental Determination of Thermal Entanglement in Spin Clusters using Magnetic Susceptibility Measurements}

\author{A. M. Souza}
\email{amsouza@cbpf.br} \affiliation{Centro Brasileiro de
Pesquisas Físicas, Rua Dr.Xavier Sigaud 150, Rio de Janeiro
22290-180, RJ, Brazil}

\author{M. S. Reis}
\affiliation{CICECO, Universidade de Aveiro, Aveiro 3819-193, Portugal}

\author{D. O. Soares-Pinto}
\affiliation{Centro Brasileiro de
Pesquisas Físicas, Rua Dr.Xavier Sigaud 150, Rio de Janeiro
22290-180, RJ, Brazil}
\affiliation{CICECO, Universidade de Aveiro, Aveiro 3819-193, Portugal}

\author{I. S. Oliveira}
\author{R. S. Sarthour}
\affiliation{Centro Brasileiro de
Pesquisas Físicas, Rua Dr.Xavier Sigaud 150, Rio de Janeiro
22290-180, RJ, Brazil}

\date{\today}

\begin{abstract}
The present work reports an experimental observation of thermal entanglement in a clusterized spin chain formed in the compound Na$_2$Cu$_5$Si$_4$O$_{14}$. The presence of entanglement was investigated through two measured quantities, an 
Entanglement Witness and the Entanglement of Formation, both derived 
from the magnetic susceptibility. It was found that pairwise entanglement exists below $ \sim 200$ K. Tripartite entanglement was also observed below  $ \sim 240$ K. A theoretical study of entanglement evolution as a function of applied field and temperature is also presented.

\end{abstract}

%\pacs{ }

\maketitle

\section{\label{sec1} Introduction}

Since the early years of quantum mechanics, entanglement has 
attracted much attention due to its
fascinating features, such as non$-$locality, as exemplified in the
Einstein, Podolsky and Rosen paradox \cite{epr}. Recently, it has 
been discovered that entangled states
constitute a valuable resource for quantum information processing
\cite{nielsen}, and it has raised a great number of studies about entanglement 
in many different quantum systems. Over the past few years, much effort has 
been done in developing methods to detect and quantify entanglement. 

Until a few years ago, entanglement was not believed to exist beyond atomic scale. The most common arguments
against entanglement on larger scales is that
macroscopic objects contain a large number of constituents 
that interact with its surroundings, inducing the decoherence phenomena which leads 
to loss of entanglement as size, complexity and system's temperature increases. Surprisingly, it was
theoretically demonstrated \cite{nielsenphd,vedral}
that entangled states can exist in solids at finite temperature and this kind of entanglement is referred in literature as \textquotedblleft thermal entanglement \textquotedblright. Since the
publication of several theoretical works 
\cite{nielsenphd,vedral,osterloh,wang2,vidal,wang3,asoudeh,wu,shawish,zhang,wang,dowling,toth,sarandy,guhne,guhne2,brukner2,wiesniak2}, a few experimental
evidences have been reported \cite{ghosh,vertesi,brukner,rappoport}, confirming the presence of
entanglement in solids state systems. 

The study of entanglement in solid state 
physics (for a detailed review see \cite{amico}) is of great relevance to the area of Quantum Information and
Quantum Computation, since many proposals of quantum chips are
solid state-based. Furthermore, the demonstration that entanglement 
can change the thermodynamical properties of solids, such as 
magnetic susceptibility \cite{ghosh}, shows that  entanglement can be related to significant 
macroscopic effects. Hence, this subject establishes a interesting 
connection between quantum information theory and condensed matter physics. 

The task of entanglement quantification is still an open problem 
in general case (for recent reviews see \cite{plenio,bruss,mintert}). Many quantities have been proposed to quantify entanglement, one 
of them is the Entanglement of Formation (EF) \cite{bennett,wootters}, which can be easily 
calculated in the case of two spin $1/2$ particles, but it can not be measured directly 
in most of the cases. Hence, usually the detection of entanglement is done 
using a quantity called Entanglement Witness (EW) \cite{Horodecki}, which is an observable that, by definition, has a 
positive expectation value for separable states and negative for some entangled states.  

In this work, we report an experimental observation of thermal entanglement in the spin chain system formed in the 
compound Na$_2$Cu$_5$Si$_4$O$_{14}$ \cite{ReisPRB,santos}, by using a thermodynamical EW based on 
the magnetic susceptibility \cite{NJP}. Furthermore, we have also derived, using the complete knowledge 
of the Hamiltonian of the spin chain, a relationship between the Entanglement of 
Formation of each pair of spins and the experimental magnetic susceptibility, which allowed to determination of  
the EF directly from the experimental data.

This paper is organized as follows: in the section \ref{sec2} we give a brief description of the particular 
system studied here. The following sections, third and fourth, 
contain the experimental results at zero applied field and a theoretical study of the entanglement 
evolution as a function of temperature and applied field respectively. In the last section, some comments and conclusions are drawn.

\section{\label{sec2} The Spin Chain Description}

\begin{figure}[tbp]
\begin{center}
\includegraphics[width=8.5cm]{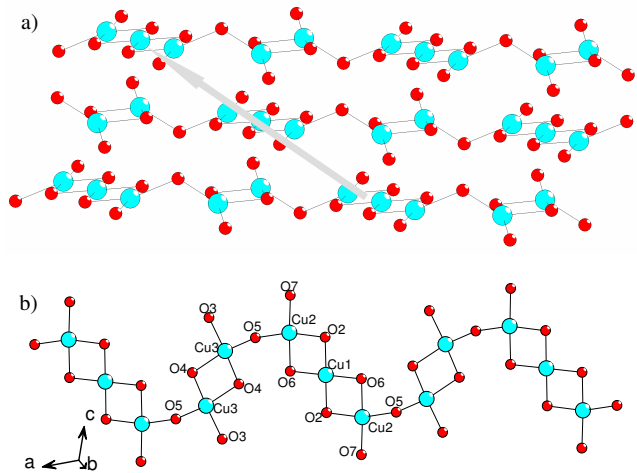}
\end{center}
\caption{The structure of the Na$_2$Cu$_5$Si$_4$O$_{14}$ compound, being the smaller circles the Oxygen atoms and the larger ones the Copper atoms, from two different views. 
(a) Side view, with an arrow indicating the staggered staking of the chains. (b) Top view of the chain.} \label{cu}
\end{figure}

\begin{figure}[tbp]
\begin{center}
\includegraphics[width=8.5cm]{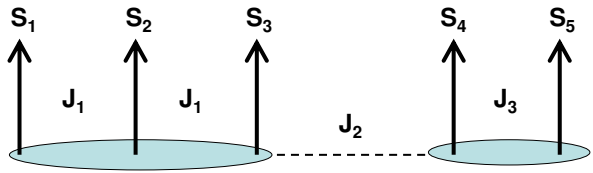}
\end{center}
\caption{(Color online) Schematic representation of the dimer-trimer cluster, and the respective 
interactions ($J_i$) between the Cu atoms, according to Equation (\ref{H}).} \label{cu2}
\end{figure}

The Copper atoms in the compound Na$_2$Cu$_5$Si$_4$O$_{14}$ have $S=1/2$ and are separated in
two groups, containing two and three atoms each, named dimer and trimer 
respectively. The whole
structure of this compound is comprised of zig-zag clusters of
copper and oxygen atoms then forming dimer-trimer sets of spins, 
as shown in Figure (\ref{cu}). There is an indirect
exchange interaction between the spins, through the electronic
oxygen clouds, and is of short range order, i.e., mainly between
the first neighbors. As previously shown \cite{ReisPRB}, the three spins that form the trimer are coupled
antiferromagnetically with each other, whereas the two atoms of the
dimer are coupled ferromagnetically. In addition, the two sets of
spins, dimer and trimer, interact antiferromagnetically with
each other. Therefore, labeling the spins according to the Figure
(\ref{cu2}), the magnetism of the this system can be described by
the Hamiltonian:

\begin{equation}
\label{H} \mathcal{H} = -J_1(\vec{S_1} \cdot \vec{S_2}  + \vec{S_2} \cdot \vec{S_3}) - J_2 (\vec{S_A}
\cdot \vec{S_B}) - J_3 (\vec{S_4} \cdot \vec{S_5}) -g\mu_B \vec{H} \cdot \vec{S}
\end{equation}
where $g$ is the Land\'e $g$-factor, $\mu_B$ is the Bohr magneton,
$\vec{H}$ is an external magnetic field, $\vec{S_A} = \vec{S_4} +  \vec{S_5}$ is the total spin of the 
dimer, $\vec{S_B} = \vec{S_1} +  \vec{S_2} + \vec{S_3}$ is the total spin of the trimer and $\vec{S} = \vec{S_A} + \vec{S_B}$ stands 
for the total spin of the cluster. The values of the exchange integrals were
determined experimentally \cite{ReisPRB}, and they were found to be  $J_1 =
-224.9$ K, $J_3 =40.22 $ K and $J_2 = -8.01 $ K, for
intra-trimer, intra-dimer and dimer-trimer, respectively. 

At finite temperatures $T$, the thermal equilibrium quantum state of the system is described by
the density matrix $\rho(T) = \exp(-\mathcal{H}/k_BT)/Z$ where $Z$ is the partition function. From $\rho(T)$ it
is possible to calculate thermodynamical quantities, such as magnetic 
susceptibility. Since the Hamiltonian $\mathcal{H}$ commutes with
the spin component along the $z$-direction $S^z$, one can show
that the magnetic susceptibility along a given direction $\alpha$ can be written as \cite{NJP}:

\begin{eqnarray}
\chi^{\alpha}(T) = \frac{(g \mu_B)^2}{k_B T} \left( \sum_{i,j=1}^{N} \langle S^{\alpha}_i S^{\alpha}_j \rangle -
 \left\langle \sum_{i=1}^{N} S^{\alpha}_i  \right\rangle^2 \right)
 \label{sus1}
\end{eqnarray}

\begin{figure}[tbp]
\begin{center}
\includegraphics[width=9.0cm]{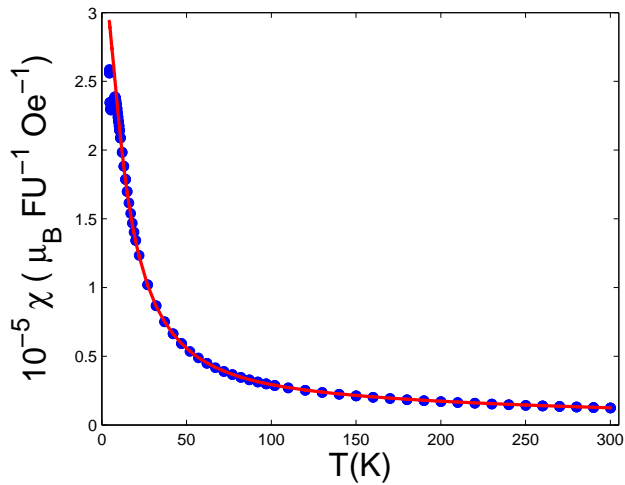}
\end{center}
\caption{ Magnetic susceptibility as a function
of temperature with an applied field of $100$ Oe. The  points ({\color{Blue}{$\bullet$}}) are 
the experimental results and 
the solid line is the theoretical prediction, based on Equation (\ref{sus1}). } \label{model}
\end{figure}

On Figure (\ref{model}), the theoretical prediction of Equation (\ref{sus1}) is compared to the experimental
magnetic susceptibility measured using a conventional superconducting quantum interference device (SQUID) magnetometer with an applied field of $100$ Oe as a function
of temperature. As can be seen from the figure, there is a monotonic increase of susceptibility down to 8 K, where there is a sharp drop associated to a transition to a 3D ground state \cite{ReisPRB}. The good agreement between the
calculations and the experimental data above 8 K shows the faithful of the dimer-trimer 
cluster model described by Hamiltonian (\ref{H}).

\section{\label{sec3} Entanglement for $H = 0$}

In this section, two quantities are used to investigate the presence of entanglement: an 
Entanglement Witness and the Entanglement of Formation, both obtained 
from the magnetic susceptibility. This study was carried out for the whole system and also 
for subsystems comprised by two and three spins.  

\subsection{Magnetic Susceptibility Witness}

The concept of Entanglement Witness was first introduced by Horodecki et al. \cite{Horodecki}, where it was defined that an EW 
is an observable which is capable to identify if a system is in an 
entangled state. However, in most of cases, an EW can not quantify the amount of entanglement present 
in the system and it gives only a sufficient condition for the presence of entanglement. In other words, an EW 
can only tell if, for some condition, a given system is in an entangled state, but if this condition 
is violated, we can not state with absolute certainty that the system's state is separable. 

Up to now, various thermodynamical 
quantities, such as magnetization \cite{brukner2}, internal energy \cite{wang,dowling,toth,sarandy,guhne,brukner2,wiesniak2} and 
heat capacity \cite{wiesniak2}, have been proposed as an EW, but these depend on the complete knowledge of the specific 
model describing the system and usually are not directly measurable quantities. However, recently, it
has been demonstrated, by Wie\'sniak et al. \cite{NJP}, that the magnetic susceptibility can be used as an EW 
for a large class of systems. This witness can be measured directly and can be applicable, 
in principle, without the full knowledge of the model Hamiltonian. If a system which 
has the symmetry $[\mathcal{H},S^z]=0$ is in an entangled state, then
the average of the magnetic susceptibility $\bar{\chi}^{exp}(T)$ measured along the three orthogonal axis satisfies the relation \cite{NJP}:

\begin{equation}
\bar{\chi}^{exp}(T) = \frac{\chi^x(T) + \chi^y(T) + \chi^z(T)}{3} < \frac{(g \mu_B)^2 N S}{3 k_B
T} \label{sus}
\end{equation}

Where $N$ is the number of spin$-S$ particles. From Equation (\ref{sus}), we can define the Entanglement
Witness as being:

\begin{equation}
EW(N)  = \frac{3 k_B T \bar{\chi}^{exp}(T)}{(g\mu_B)^2 N S} - 1 \label{wit}
\end{equation}

Thus, systems with $EW(N) < 0$ are in an entangled state. However, it must be emphasized that the
condition $EW(N) \geq 0 $ does not necessarily implies in separability. With $EW(5)$ obtained 
from the measured magnetic susceptibility, we can determine the presence of entanglement in our 5 spins system. However 
it is also interesting to study if the subsystems are in an entangled state. In order to analyze the entanglement between spins in a subsystem, it is necessary to derive the
contribution of the susceptibility $\bar{\chi}_{sub}(T)$ due to only the spins that matter. Theoretically, this 
can be done according to Equation (\ref{sus1}), but using the reduced density matrix $\rho_{sub}(T)$, which represents the
density matrix of the subsystem. The $\rho_{sub}(T)$ can be obtained 
by using the partial trace operation \cite{nielsen}, which sums over all the possible states 
of the spins in the system, except those belonging to the subsystem. From the numerical calculation 
of the total susceptibility and subsystem's contribution, we can define the ratio
$R_{sub}^{theo}(T) = \bar{\chi}_{sub}^{theo}(T)/ \bar{\chi}^{theo}(T)$ which represents the fraction of the subsystem's contribution 
to the total magnetic susceptibility. With this quantity at hand, which can only be calculated with the knowledge of the Hamiltonian, it is possible 
to extract separately, from the experimental data, that contains the contribution of all spins, the magnetic susceptibility of the trimer  $\bar{\chi}_{Tri}^{exp}(T) = R_{Tri}^{theo}(T) \times \bar{\chi}^{exp}(T)$, dimer $\bar{\chi}_{Dim}^{exp}(T) = R_{Dim}^{theo}(T) \times \bar{\chi}^{exp}(T)$ and so on.

\begin{figure}[tbp]
\begin{center}
\includegraphics[width=9.0cm]{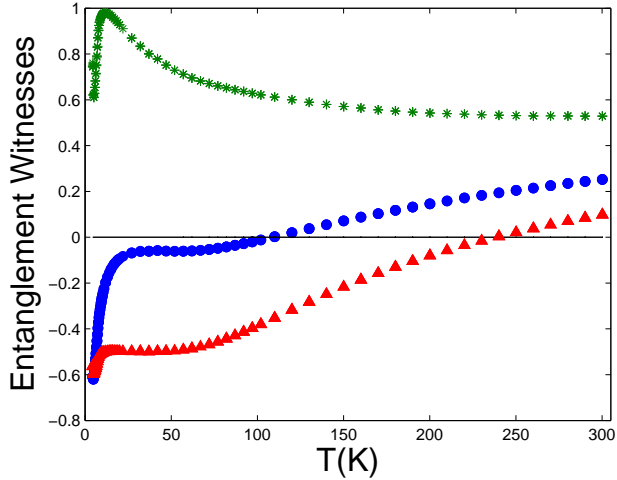}
\end{center}
\caption{ Entanglement Witness for the total system $EW(5)$ circles ({\color{Blue}{$\bullet$}}),  for the 
trimer $EW(3)$  triangles ({\color{red}{$\blacktriangle$}}) and for the dimer $EW(2)$ stars ({\color{Green}{*}}).} \label{chiwit} 
\end{figure}

In Figure (\ref{chiwit}), we show the experimental data, as can be seen, the $EW(5)$ extracted from the total magnetic susceptibility 
is negative for any temperature below $\sim 110 $ K, showing the presence of entanglement in the system. Furthermore, the fact that $EW(3)$ for the trimer is negative below $\sim 240 $ K and $EW(2)$ for the dimer is always positive suggests that entanglement only occurs between spins inside the trimer.

\subsection{Entanglement of Formation}

The Entanglement of Formation \cite{bennett,wootters}, which can be seen as the amount of quantum resources needed to create 
a given entangled state, is one of the most used for entanglement measurement. This was proposed to quantify the entanglement of a bipartite system. Unfortunately, as many others proposals, the EF is extremely difficult to calculate in general. However, in the special case of two spin $1/2$ particles, recently an analytical expression was 
derived by Wootters who showed that the EF of any density matrix ($\rho$) of two spin $1/2$ particles is given by
\cite{wootters}:

\begin{equation}
EF = - x log_2(x) - (1-x) log_2(1-x) \label{ef}
\end{equation}
where $x = ( 1+ \sqrt{1-C^2} )/2$, being $C$ the concurrence,
defined as $ \max(0, \sqrt{\Lambda_1} -  \sqrt{\Lambda_2} -
\sqrt{\Lambda_3}  - \sqrt{\Lambda_4})$ and $\Lambda$'s are the
eigenvalues of $R = \rho_{ } \sigma_y \otimes \sigma_y \rho^*
\sigma_y \otimes \sigma_y$, labeled in decreasing order. The degree of entanglement varies from $0$ to $1$, a pair of spins is 
considered to be in a maximally entangled state if $EF = 1$ and separable when EF is equal to zero, for any other values the state of the spins is said to be partially entangled. 

Since the Hamiltonian (\ref{H}) commutes with $S^z$, we can write
the reduced density matrix of the spins located in the sites $i$
and $j$ as \cite{ring,wang}:

\begin{equation}
\rho_{ij}(T) = \left(
\begin{array}{cccc}
u^+ & 0 & 0 & 0\\
0 & \omega_1 & z ^*& 0\\
0 & z & \omega_1 & 0\\
0 & 0 & 0 & u^-
\end{array}
\right) \label{rij}
\end{equation}
being $u^\pm = (1 \pm 2\langle S^z_i + S^z_j \rangle + 4\langle
S^z_i S^z_j \rangle )/4 $ and $z = \langle S^x_i S^x_j \rangle +
\langle S^y_i S^y_j \rangle  + i \langle S^x_i S^y_j \rangle  - i
\langle S^y_i S^x_j \rangle $. The concurrence of such density
matrix can be written as:

\begin{equation}
C = 2 \max (0,|z| - \sqrt{u^+u^-} ) \label{c}
\end{equation}

Now, exploring the fact that the system is isotropic at zero magnetic field, it is possible to
write $\langle S^x_i S^x_j \rangle = \langle S^y_i S^y_j \rangle =
\langle S^z_i S^z_j \rangle = G_{ij}/3$ and $ \langle S^x_i
S^y_j \rangle = \langle S^y_i S^x_j \rangle $. Then,  one can
rewrite the concurrence between spins $i$ and $j$ in terms of the
correlations functions, such as $C_{ij}(T) = \frac{2}{3} \max(0,2|G_{ij}|- G_{ij}-\frac{3}{4})$. It is straightforward to
verify from (\ref{sus1}) that $\bar{\chi}_{ij}^{exp}(T) = 2 (g \mu_B)^2 (1/4 + G_{ij}/3)/k_BT $ and thus
the concurrence becomes:

\begin{equation}
C_{ij}(T)= \frac{k_B T}{(g\mu_B)^2} \max \left(0, 2 \left| \bar{\chi}_{ij}^{exp}(T)-\frac{(g\mu_B)^2}{2 k_B T}\right| - \bar{\chi}_{ij}^{exp}(T)\right)
 \label{c2}
\end{equation}

The Equation (\ref{c2}) relates the concurrence of the pair $i-j$, hence the EF, to the magnetic susceptibility $\bar{\chi}_{ij}^{exp}(T)$, which can be obtained from the experimental data as explained before. It is interesting to note that the Equation (\ref{c2}) leads to the concurrence found by Asoudeh and Karimipour \cite{asoudeh} for mean-field clusters. 

\begin{figure}[tbp]
\begin{center}
\includegraphics[width=9.0cm]{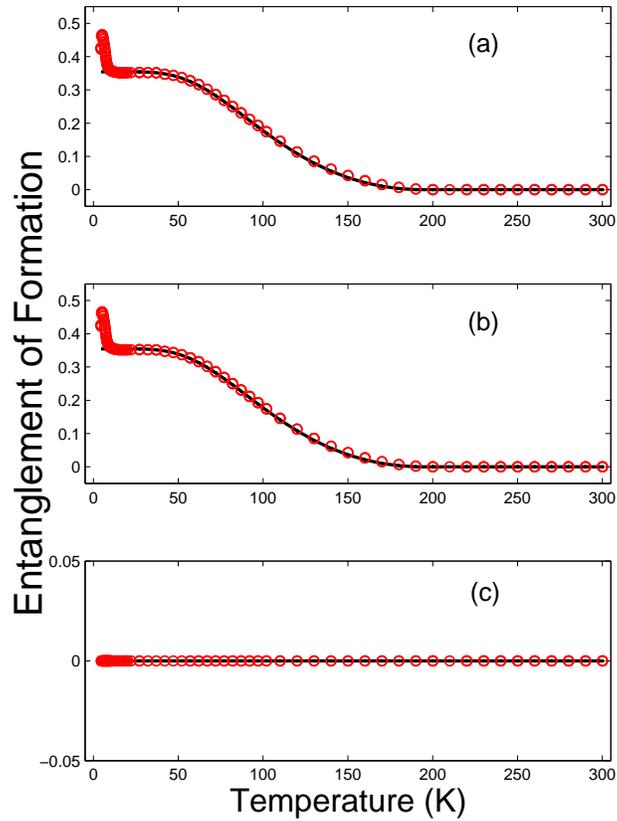}
\end{center}
\caption{  The experimentally determined Entanglement of Formation, the circles, for 
the pairs inside the trimer $1-2$ (a), $2-3$ (b) and $1-3$ (c). The solid lines are the theoretical prediction. The small discrepancy between theory and experiment at low temperatures is associated to a 3D transition \cite{ReisPRB}.} 
\label{Ef}
\end{figure}

In Figure (\ref{Ef}) we show the experimental EF obtained from the measured magnetic susceptibility and the theoretical EF calculated with the reduced density matrix $\rho_{ij}(T)$ for the pairs inside the trimer ($1-2$,$2-3$ and $1-3$). We can see that there is
entanglement only between the pairs of spins $1-2$ and $2-3$ which persists up to a critical temperature $ T^c \sim 200$ K. The Entanglement of Formation for other pairs are always null and they are not shown. These results confirm that only the spins in the trimer are entangled, being those in the dimer in a separable state. A interesting feature is that the Entanglement Witness for the trimer $EW(3)$ gives $T^c \sim 240$ while the $T^c$ obtained from the Entanglement of Formation is $\sim 200$ K. Since the EF can only see bipartite entanglement, this result indicates that pairwise
entanglement is not the sole entanglement in the system and a tripartite entanglement is also present in the trimer.  

It is also interesting to ask if this multipartite entanglement is genuine or not. Genuine multipartite 
entanglement can be identified by the criterion described in \cite{guhne2}. Applied to the 
trimer system, the criterion states that if  the expression $ \langle \mathcal{H}_{Tri} \rangle < J_1 (1 + \sqrt{5})/4 $ holds, where 
$ \langle \mathcal{H}_{Tri} \rangle $ is the mean value of the energy corresponding just to the trimer part,  then the 
tripartite entanglement is indeed genuine. We have numerically calculated, from the 
reduced density matrix of the trimer, the threshold temperature below which genuine tripartite 
entanglement exists. We found that the threshold temperature is $ \sim 108 $ K. This result suggests 
that the compound has a genuine tripartite entanglement for a finite temperature. Therefore, this genuine 
entanglement does not extend up to $ 240 $ K. It is also important to emphasize the fact that 
entanglement is confined in the trimer and thus, it is not a macroscopic entanglement, but rather 
a thermal entanglement.      

\section{\label{sec4} The effect of the magnetic field}

\begin{figure}[tbp]
\begin{center}
\includegraphics[width=14.0cm]{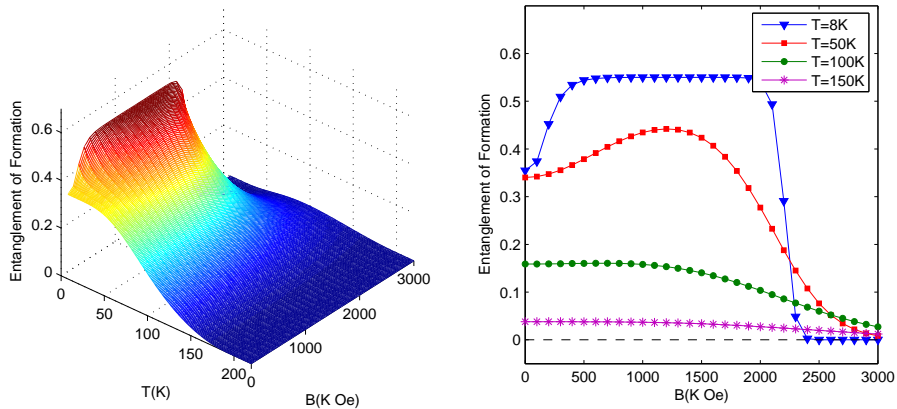}
\end{center}
\caption{(Left) Entanglement of Formation (EF) for the pairs $1-2$ and $2-3$, as a function of temperature and applied field in the H-T plane. (Right) EF for selected temperatures, as a function of applied magnetic field .} \label{b12}
\end{figure}

\begin{figure}[tbp]
\begin{center}
\includegraphics[width=14.0cm]{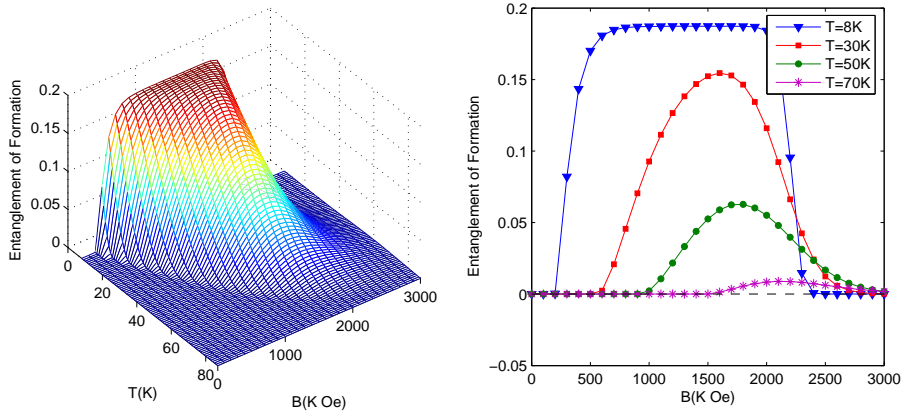}
\end{center}
\caption{(Left) Entanglement of Formation (EF) for the pair $1-3$, as a function of temperature and applied field in the H-T plane. (Right) EF for selected temperatures, as a function of applied magnetic field.} \label{b13}
\end{figure}

In order to investigate the effect of the application of a magnetic field,
the EF was calculated for each pair of
spins. In Figures (\ref{b12}) and  (\ref{b13}) we show the EF for the
$3$ pairs of spins in the trimer. The entanglement of the pair
$1-2$ behaves as the same way as of the pair $2-3$, the application of a low
magnetic field increases the amount of
entanglement of both pairs at low temperatures and also creates a small
entanglement between the pair $1-3$, which was zero at zero field. Similar behavior was 
also encountered in other theoretical works \cite{vedral,asoudeh} and can be understood in 
terms of changes in the ground state. At $H = 0$, the ground state
of the system is degenerated, being a statistical mixture of two states. This mixture has only 
entanglement between pairs $1-2$ and $2-3$. The application of an external magnetic field changes the
energy eigenvalues leading to a different ground state which has a different degree of entanglement. 

For a high enough magnetic field, the ground sate becomes the state $\mid \uparrow \uparrow \uparrow \uparrow \uparrow \rangle$, which correspond to all spins aligned parallel with the external field. At low temperatures, where the ground state is very populated, EF vanishes suddenly near $\sim 2200$ Oe 
due to the sudden change in the ground state to  $\mid \uparrow \uparrow \uparrow \uparrow \uparrow \rangle$. However for higher temperatures, there are many states populated and the decreasing of entanglement is slow. 

From the $H-T$ diagram showed in Figures (\ref{b12}) and (\ref{b13}), we can also observe that the entanglement of pairs do not occurs above the critical temperature $T^{c} \sim 200$ K and above the critical field  $H^{c} \sim 3000$ Oe. A interesting feature is 
that the critical field $H^c$ is much higher than the available field intensities that usually can be create in laboratories, which indicates that the entanglement in this system can not be destroyed by common magnetic field for temperatures below $200$ K, providing that high magnetic fields do not induce any structural changes in the compound. The Entanglement of Formation for others pairs are always zero at any point in the $H-T$ plane and they are not shown.

\section{\label{sec5} Conclusions}

In summary, we have successfully established the presence of thermal entanglement in the compound Na$_2$Cu$_5$Si$_4$O$_{14}$. From the obtained
results, we could conclude that entanglement is strong in this
system and disappears only at high temperatures and high magnetic fields. At zero magnetic field we found experimentally that pairwise entanglement exists below $ \sim 200$ K and there is tripartite entanglement below $ \sim 240$ K, which shows that multipartite entanglement is stronger than bipartite entanglement in our case. Same feature was also verified experimentally in the compound Na$_2$V$_3$O$_7$ \cite{vertesi}. The entanglement observed here is confined into small cluster of spin and thus it is not a macroscopic entanglement. We have also presented a theoretical study of the entanglement evolution as a function of applied field and temperature, showing that magnetic field can increase the degree of entanglement in the pairs $1-2$, $2-3$ and $1-3$.

We emphasize that some results of this paper is based on the validity of the dimer-trimer model, however this model is shown to be a good model for the present system (see Figure (\ref{model})). Furthermore, since the Entanglement Witness $EW(5)$ does not require any assumption
about the model, or the explicit values of the exchange-coupling parameters, if the model is not completely correct, the main conclusion does not change, i.e the presence of thermal entanglement in the compound Na$_2$Cu$_5$Si$_4$O$_{14}$.

\begin{acknowledgments}
The authors acknowledge support from the Brazilian funding
agencies CNPq, CAPES and the Brazilian
Millennium Institute for Quantum Information. The authors also
acknowledge J. Rocha, P. Brandão and A. Moreira dos Santos for their strong contribution 
to the full characterization of the compound from the crystallographic and magnetic point of view. DOSP would like to thanks CAPES for the financial support at Universidade de Aveiro at Portugal.
\end{acknowledgments}

\bibliography{EntCluster}

\end{document}